\documentclass[aps,prb,twocolumn,superscriptaddress,floatfix]{revtex4-1}

\usepackage{graphicx,graphics}
\usepackage{dcolumn}
\usepackage{amsmath,amssymb,amsfonts}
\usepackage{latexsym,verbatim}
\usepackage{bm,bbold}
\usepackage{color}
\usepackage{ulem}
\usepackage[percent]{overpic}
\usepackage[breaklinks=true,colorlinks,citecolor=blue,linkcolor=blue,urlcolor=blue]{hyperref}
\begin{document}
\title{Theory of third harmonic generation in graphene: a diagrammatic approach}
\author{Habib Rostami}
\email{Habib.Rostami@iit.it}
\affiliation{Istituto Italiano di Tecnologia, Graphene Labs, Via Morego 30, I-16163 Genova,~Italy}
\affiliation{NEST, Istituto Nanoscienze-CNR and Scuola Normale Superiore, I-56126 Pisa,~Italy}
\author{Marco Polini}
\affiliation{Istituto Italiano di Tecnologia, Graphene Labs, Via Morego 30, I-16163 Genova,~Italy}
\begin{abstract}
We present a finite-temperature diagrammatic perturbation theory of third harmonic generation (THG) in doped graphene. 
We carry out calculations of the third-order conductivity in the scalar potential gauge, highlighting a subtle cancellation between a Fermi surface contribution, which contains only power laws, and power-law contributions of inter-band nature. Only logarithms survive in the final result. We conclude by presenting quantitative results for the up-conversion efficiency at zero and finite temperature. Our results shed light on the on-going dispute over the dependence of THG on carrier concentration in graphene.
\end{abstract}

\maketitle

{\it Introduction.---}The non-linear optical properties of graphene~\cite{geim_naturemater_2007}, the most studied two-dimensional (2D) material, are beginning to attract considerable interest. Using four-wave mixing, Hendry et al.~\cite{hendry_prl_2010} demonstrated experimentally that the third-order optical susceptibility of graphene is remarkably large ($\approx 1.4\times10^{-15} ~{\rm m}^2/{\rm V}^2$) and only weakly dependent on wavelength in the near-infrared frequency range. Third harmonic generation (THG) from mechanically exfoliated graphene sheets has been measured by Kumar et al.~\cite{kumar_prb_2013} who extracted a value of the third-order susceptibility on the order of $10^{-16}~{\rm m}^2/{\rm V}^2$ for an incident photon energy $\hbar\omega=0.72~{\rm eV}$. Finally, Hong et al.~\cite{hong_prx_2013} reported strong THG in graphene grown by chemical vapor deposition, in the situation in which the incident photon energy $\hbar\omega=1.57~{\rm eV}$ is in three-photon resonance with the exciton-shifted van Hove singularity.

\begin{figure}[h!]
\centering
\begin{overpic}[width=0.8\linewidth]{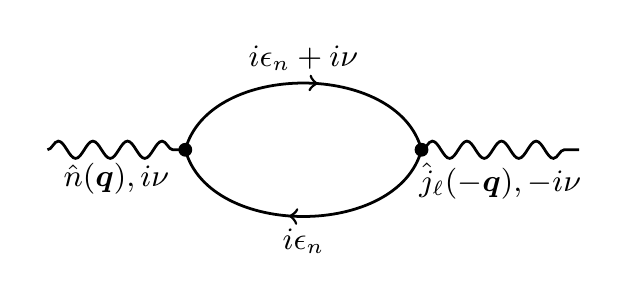}\put(0.0,40){(a)}\end{overpic}
\begin{overpic}[width=0.8\linewidth]{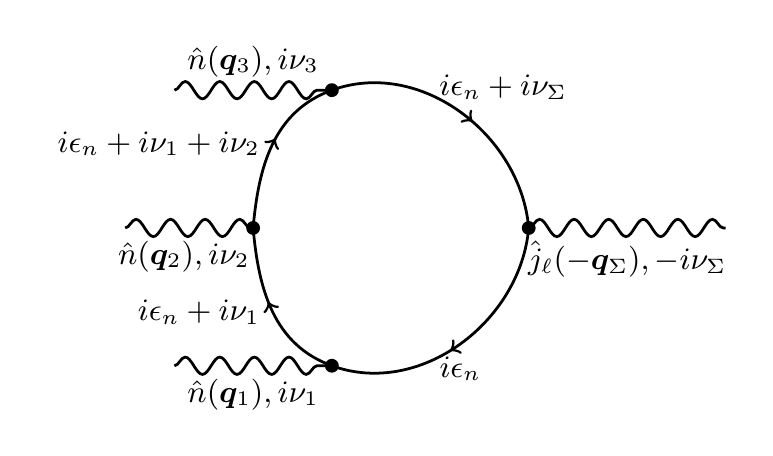}\put(0.0,55){(b)}\end{overpic}
\caption{\label{fig:one}Panel (a) ``Bubble'' diagram for the first-order current response to a scalar potential. Panel (b) Example of a four-leg  diagram for the third-order response tensor $\Pi^{(3)}_{\ell}(-\nu_\Sigma;\nu_1,\nu_2,\nu_3|-{\bm q}_\Sigma;{\bm q}_1,{\bm q}_2,{\bm q}_3)$ in the SPG. Solid lines indicate non-interacting Matsubara Green's functions. Wavy lines on the left side of the loop indicate scalar potentials (incoming photons) carrying a finite wave vector ${\bm q}_i$ and energy $\nu_i$ with $i=1\dots3$. The wavy line on the right of the fermion loop denotes a scalar potential (outgoing photon) carrying a wave vector ${\bm q}_{\Sigma}$ and energy $\nu_{\Sigma}$. Conservation of momentum and energy require ${\bm q}_{\Sigma} = \sum_i {\bm q}_i$ and $\nu_{\Sigma}=\sum_i \nu_i$, respectively. Black dots indicate external vertices. Here $\nu_i$ ($\epsilon_n$) denotes a bosonic (fermionic) Matsubara energy.}
\end{figure}

Despite the large body of theoretical work on THG in graphene~\cite{jafari_jpcm_2012,mikhailov_prb_2014,mikhailov_arxiv_2015,
cheng_njp_2014,cheng_prb_2015}, no consensus appears to exist among different authors (or, for that matter, even between different articles of the same author~\cite{mikhailov_prb_2014,mikhailov_arxiv_2015}). In this work we present a finite-temperature diagrammatic perturbation theory of THG in graphene (see Fig.~\ref{fig:one}). Our approach has the advantage of being transparent and easily extendable to 2D materials with a more complex band structure like graphene derivatives (e.g.~bilayer graphene), transition-metal dichalcogenides, and few-layer black-phosphorus. Also, it can be generalized~\cite{yu_prb_1991,jafari_jpsj_2006,katsnelson_jpcm_2010} to take into account electron-electron interactions (plasmons, excitons, etc.). 

We carry out microscopic calculations of THG in a non-interacting 2D system of massless Dirac fermions (MDFs)~\cite{kotov_rmp_2012} in the {\it scalar potential gauge} (SPG). In this gauge, light-matter interactions are described by utilizing an external scalar potential, which couples to the electronic density operator. As explained in Ref.~\onlinecite{abedinpour_prb_2011}, this gauge is free of the pathologies that one encounters when optical properties of 2D MDFs are calculated by employing the {\it vector potential gauge} and the MDF current operator, which lacks a diamagnetic contribution~\cite{chirolli_prl_2012}. Furthermore, in the vector potential gauge, light-matter interactions are described through the minimal coupling ${\bm p} \rightarrow {\bm p} + e {\bm A}(t)/c$ in the continuum-model Hamiltonian. The vector potential is time-dependent but uniform, implying that in this gauge the momentum $\hbar {\bm q}$ of incident photons is set to zero from the very beginning. Although this choice tremendously simplifies analytical calculations, it is known to miss intra-band (i.e.~Fermi surface) contributions. 

In this work we report an {\it a priori} unexpected cancellation between the Fermi surface contribution to the third-order conductivity, which contains only power laws, and power-law contributions of inter-band nature. Only logarithms survive in the final result. This anomalous cancellation occurs for all values of the microscopic parameters and is not tied to the linear dispersion of MDFs in single-layer graphene. For example, we have checked (not shown here) that it also occurs for massive chiral fermions in bilayer graphene~\cite{geim_naturemater_2007}. Also, we have checked (not shown here) that it occurs i) in the presence of terms that break particle-hole symmetry (e.g.~next-nearest-neighbor hopping in the tight-binding model) and ii) for anisotropic 2D MDFs (e.g.~uniaxially strained graphene).  We believe that this cancellation stems from the gapless nature of the dispersion relation, as power-law terms are present in the final result for THG in gapped graphene~\cite{jafari_jpcm_2012}. (However, the calculations of Ref.~\onlinecite{jafari_jpcm_2012} are in the vector potential gauge.) Our final result---see Eq.~(\ref{eq:final_result}) below---agrees with the clean-limit result of Refs.~\onlinecite{cheng_njp_2014,cheng_prb_2015}, and disagrees with the clean-limit result of {\it both} Refs.~\onlinecite{mikhailov_prb_2014,mikhailov_arxiv_2015}.

{\it Diagrammatic perturbation theory of THG.---}We consider the single-channel Hamiltonian of a 2D system of non-interacting MDFs~\cite{kotov_rmp_2012}, $\hat{\cal H}_0  = v_{\rm F} \int d^2{\bm r}~\hat{\psi}^\dagger({\bm r})({\bm \sigma} \cdot {\bm p})
\hat{\psi}({\bm r})$, where $v_{\rm F}\sim 10^{6}~{\rm m}/{\rm s}$ is the graphene Fermi velocity, $\hat{\psi}({\bm r}) = (\hat{\psi}_{\rm A}({\bm r}),\hat{\psi}_{\rm B}({\bm r}))^{\rm T}$, ${\bm \sigma}=(\sigma_x,\sigma_y)$ is a 2D vector of Pauli matrices, and ${\bm p} = -i\hbar \nabla_{\bm r}$. 
We calculate THG by using perturbation theory in an external {\it homogeneous} time-dependent electric field ${\bm E}(t)$. The latter induces a current, which can be formally expanded in powers of the electric field: $J_{\ell} =\sum_n J^{(n)}_{\ell}$, where $n=1,2,3,\dots$ denotes the order in perturbation theory and $\ell=x,y$ is a Cartesian index. Due to spatial inversion symmetry, the second-order ($n=2$) response to a uniform electric field is identically zero~\cite{Landau08,Boyd}. 
The first- and third-order conductivity tensors, ${\bm \sigma}^{(1)}$ and ${\bm \sigma}^{(3)}$, are defined as following: 
\begin{equation}\label{eq:conductivity1}
J^{(1)}_{\ell}(\omega) = \sum_{\alpha_1}\sigma^{(1)}_{\ell \alpha_1}(-\omega;\omega)E_{\alpha_1}(\omega)
\end{equation}
and
\begin{equation}\label{eq:conductivity3}
J^{(3)}_{\ell}(\omega_\Sigma) = \sum_{\alpha_{1}\dots\alpha_{3}}\sigma^{(3)}_{\ell \alpha_1 \alpha_2 \alpha_3}(-\omega_\Sigma;\omega_1,\omega_2,\omega_3) 
\Pi^{3}_{i=1} E_{\alpha_i}(\omega_i)~,
\end{equation}
where $\alpha_1\dots \alpha_n =x,y$ are Cartesian indices, $E_{\alpha_i}$ denotes the $\alpha_i$-th Cartesian component of ${\bm E}$, and $\omega_\Sigma=\sum_i\omega_i$. The quantity $\sigma^{(1)}_{\ell \alpha_1}(-\omega;\omega)$ denotes the Cartesian components of the usual linear-response conductivity tensor~\cite{basov_rmp_2014}.

Coupling of the electronic degrees of freedom described by $\hat{\cal H}_0$ to the electric field of incident light can be described in different electromagnetic gauges~\cite{Giuliani_and_Vignale}. In the SPG, light-matter interactions are described by adding a scalar potential to the Hamiltonian, i.e.~$\hat{\cal H}_V = \hat{\cal H}_0+\int d^2{\bm r} V({\bm r},t){\hat n}({\bm r})$. Here, ${\hat n}({\bm r})$ is the density operator and $V({\bm r},t) = -e \Phi({\bm r},t)$, where $\Phi({\bm r},t) = {\cal S}^{-1} [\varphi({\bm q},\omega) e^{i({\bm q}\cdot {\bm r} 
-\omega t)}e^{\eta t /\hbar} +{\rm c.c.}]/2$ is the electric potential and ${\cal S}$ the 2D electron system area. 
The quantity $\eta$ is the usual positive infinitesimal~\cite{Giuliani_and_Vignale}, which is needed to make sure that the field vanishes in the remote past ($t \rightarrow -\infty$). 
The Fourier components of the electric field are given by ${\bm E}({\bm q},\omega)= -i{\bm q} \varphi({\bm q},\omega)$. In order to have a finite electric field, the photon wave vector ${\bm q}$ must be kept finite in this gauge. The uniform $|{\bm q}|\to 0$ limit can be taken only at the end of the calculation. In the SPG we are therefore able to take into account {\it both} intra and  inter-band contributions to optical response tensors.

In the SPG, the third-order conductivity tensor ${\bm \sigma}^{(3)}$ can be obtained from
\begin{equation}\label{eq:sigmaPi3}
\sigma^{(3)}_{\ell \alpha_1 \alpha_2 \alpha_3}=  \frac{(-i)^3 (-e)^3}{{\cal N}!}\frac{\partial ^3 \Pi^{(3)}_{\ell}}
{\partial q_{1,\alpha_1}\partial q_{2,\alpha_2}\partial q_{3,\alpha_3}}
\Big |_{\{{\bm q}_i \to  {\bm 0}\}}~,
\end{equation}
where ${\cal N}!$ originates from the Taylor expansion in powers of ${\bm q}_i$.
The rank-$1$ tensor $\Pi^{(3)}_{\ell}(-\nu_\Sigma;\nu_1,\nu_2,\nu_3 ~ | -{\bm q}_\Sigma;{\bm q}_1,{\bm q}_2,{\bm q}_3)$ is a sum of Feynman diagrams like the one in Fig.~\ref{fig:one}(b):
\begin{eqnarray}\label{eq:pi3}
\Pi^{(3)}_{\ell} &=&\frac{e v_{\rm F}N_{\rm f} }{3!}\int \frac{d^2{\bm k}}{(2\pi)^2}\sum_{\lambda_1\dots \lambda_4= \pm} 
\sum_{\cal P}
F_{\lambda_1\dots \lambda_4}({\bm k}, {\bm q}_1, {\bm q}_2, {\bm q}_3)\nonumber\\
&\times&I_{\lambda_1\dots \lambda_4}({\bm k},{\bm q}_1, {\bm q}_2, {\bm q}_3,\nu_1,\nu_2,\nu_3)~,
\end{eqnarray}
where
\begin{eqnarray}\label{eq:SPG_formfactor}
F_{\lambda_1\dots \lambda_4}&=&
\big \langle\lambda_1, {\bm k} \big | \hat{n}({\bm q}_1) \big |\lambda_2, {\bm k}+{\bm q}_1 \big \rangle
\nonumber\\
& \times&
\big \langle\lambda_2, {\bm k}+{\bm q}_1 \big | \hat{n}({\bm q}_2) \big |\lambda_3, {\bm k}+{\bm q}_1+{\bm q}_2 \big  \rangle
\nonumber\\
&\times&
\big  \langle\lambda_3, {\bm k}+{\bm q}_1+{\bm q}_2 \big | \hat{n}({\bm q}_3) \big |\lambda_4, {\bm k}+{\bm q}_\Sigma \big \rangle
\nonumber\\
& \times&
\big \langle \lambda_4, {\bm k}+{\bm q}_\Sigma \big | \frac{\hat{j}_{\ell}(-{\bm q}_\Sigma)}{ -e v_{\rm F}} \big |\lambda_1, {\bm k} \big \rangle
\end{eqnarray}
is a dimensionless form factor due to the four external vertices, while
\begin{eqnarray}\label{eq:Matsubara_sum}
&&I_{\lambda_1\dots \lambda_4}
=\frac{1}{\beta}\sum_{i\epsilon_n} 
\Big[
G (i\epsilon_n, \varepsilon_{\lambda_1,{\bm k}}) 
\nonumber\\ 
&\times&
G (i\epsilon_n+i\nu_1, \varepsilon_{\lambda_2,{\bm k}+{\bm q}_1}) 
G (i\epsilon_n+i\nu_1+i\nu_2, \varepsilon_{\lambda_3,{\bm k}+{\bm q}_1+{\bm q}_2}) 
\nonumber\\
&\times&
G (i\epsilon_n+i\nu_\Sigma, \varepsilon_{\lambda_4,{\bm k}+{\bm q}_\Sigma})
\Big]
\end{eqnarray}
is due to the presence of four Green's functions in Fig.~\ref{fig:one}(b). In Eq.~(\ref{eq:pi3}), $N_{\rm f}=4$ is the number of fermion flavors in graphene~\cite{kotov_rmp_2012} and $\sum_{\cal P}$ denotes a sum over the $3! = 6$ permutations of the energy and wave vector variables $\{\nu_i,{\bm q}_i\}$ of the three incoming photons~\cite{Landau08,Boyd} in Fig.~\ref{fig:one}. In Eq.~(\ref{eq:SPG_formfactor}), $\hat{n}({\bm q})$ and $\hat{\bm j}({\bm q})$ are the Fourier transforms of the density $\hat{n}({\bm r})$ and paramagnetic current $\hat{\bm j}({\bm r})$ operators, respectively, where $\hat{\bm j}({\bm r}) = -e v_{\rm F} \hat{\psi}^\dagger({\bm r}){\bm \sigma}\hat{\psi}({\bm r})$, $-e$ being the electron charge. Since there is no vector potential in the SPG, we do not need to worry about diamagnetic contributions~\cite{chirolli_prl_2012} to the paramagnetic current operator $\hat{\bm j}({\bm r})$. In Eq.~(\ref{eq:Matsubara_sum}), $\beta=1/(k_{\rm B}T)$ where $T$ is temperature, 
$\epsilon_n = (2n+1)/\beta$ is a fermionic Matsubara energy, and $G(i\epsilon_n,\varepsilon_{\lambda,{\bm k}})=1/(i\epsilon_n -\varepsilon_{\lambda,{\bm k}})$ is the bare Green's function in the band representation, with $\varepsilon_{\lambda, {\bm k}}=\lambda \hbar v_{\rm F} |{\bm k}| $ for conduction ($\lambda=+$) and valence ($\lambda=-$) band states.  

To make progress, we must {\it first} perform the sum over the fermionic Matsubara energy $\epsilon_n$ in Eq.~(\ref{eq:Matsubara_sum}). This can actually  be done analytically by following standard textbook tricks~\cite{Giuliani_and_Vignale}.  Only {\it after}, one can carry out the analytical continuation $i\nu_i \to \hbar \omega_i +i \eta $ to real photon energies $\hbar\omega_i$.

The end result of this procedure for the case of harmonic generation~\cite{Landau08,Boyd}, i.e.~$\omega_i=\omega$ and ${\bm q}_i={\bm q}$, and the $\ell= x$ component of $\Pi^{(3)}_{\ell}$ is:
\begin{widetext}
\begin{eqnarray}\label{eq:pi32}
&&\Pi^{(3)}_{x}(-3\omega;\omega,\omega,\omega ~ | -3{\bm q};{\bm q},{\bm q},{\bm q})
 =
 N_{\rm f} e v_{\rm F}\int\frac{d^2{\bm k}}{(2\pi)^2}
\sum_{\lambda_1\dots \lambda_4 = \pm} 
\frac{F_{\lambda_1\dots\lambda_4}({\bm k},{\bm q},{\bm q},{\bm q})}
{
3(\hbar \omega+i\eta) + \varepsilon_{\lambda_1, {\bm k}} - \varepsilon_{\lambda_4, {\bm k}+3{\bm q}}
}\times
\nonumber\\
\Bigg\{
&&
\left[ 
\frac{1}{2(\hbar \omega+i\eta) + \varepsilon_{\lambda_1, {\bm k}} - \varepsilon_{\lambda_3, {\bm k}+2{\bm q}}} 
\left\{ 
\frac{n_{\rm F}(\varepsilon_{\lambda_1, {\bm k}})-n_{\rm F}(\varepsilon_{\lambda_2, {\bm k}+{\bm q}})}
{\hbar\omega + \varepsilon_{\lambda_1, {\bm k}} - \varepsilon_{\lambda_2, {\bm k}+{\bm q}}+i\eta}
-\frac{n_{\rm F}(\varepsilon_{\lambda_2, {\bm k}+{\bm q}}) - n_{\rm F}(\varepsilon_{\lambda_3, {\bm k}+2{\bm q}})}
{\hbar\omega + \varepsilon_{\lambda_2, {\bm k}+{\bm q}} - \varepsilon_{\lambda_3, {\bm k}+2{\bm q}}+i\eta}
 \right\}
 \right]
 +
 \nonumber\\&&
 \left[ 
 \frac{1}{2(\hbar \omega+i\eta) + \varepsilon_{\lambda_2, {\bm k}+{\bm q}} - \varepsilon_{\lambda_4, {\bm k}+3{\bm q}} } 
\left\{
 \frac{n_{\rm F}(\varepsilon_{\lambda_3, {\bm k}+2{\bm q}}) - n_{\rm F}(\varepsilon_{\lambda_4, {\bm k}+3{\bm q}})}
{\hbar\omega + \varepsilon_{\lambda_3, {\bm k}+2{\bm q}} - \varepsilon_{\lambda_4, {\bm k}+3{\bm q}}+i\eta}
-\frac{n_{\rm F}(\varepsilon_{\lambda_2, {\bm k}+{\bm q}}) - n_{\rm F}(\varepsilon_{\lambda_3, {\bm k}+2{\bm q}})}
{\hbar\omega + \varepsilon_{\lambda_2, {\bm k}+{\bm q}} - \varepsilon_{\lambda_3, {\bm k}+2{\bm q}}+i\eta}
 \right\}
 \right]
 \Bigg\}~.
\end{eqnarray}
\end{widetext}
In Eq.~(\ref{eq:pi32}), 
\begin{eqnarray}\label{eq:form_factor}
F_{\lambda_1\dots\lambda_4}({\bm k},{\bm q},{\bm q},{\bm q})&=&
\frac{1+\lambda_1 \lambda_2 e^{i[\phi({\bm k}+{\bm q})-\phi({\bm k})]}}{2}
\nonumber\\
&\times&
\frac{1+\lambda_2 \lambda_3 e^{i[\phi({\bm k}+2{\bm q})-\phi({\bm k}+{\bm q})]}}{2}
\nonumber\\
&\times&
\frac{1+\lambda_3 \lambda_4 e^{i[\phi({\bm k}+3{\bm q})-\phi({\bm k}+2{\bm q})]}}{2}
\nonumber\\
&\times&
\frac{\lambda_1 e^{i\phi({\bm k})} +\lambda_4  e^{-i\phi({\bm k}+3{\bm q})}}{2}
\end{eqnarray}
is the form factor in the SPG, $\phi({\bm k})$ being the polar angle of ${\bm k}$, while $n_{\rm F}(E)= \{\exp[\beta(E-\mu)] + 1 \}^{-1}$ is the usual Fermi-Dirac distribution function, $\mu$ being the finite-$T$ chemical potential. THG in graphene is therefore the result of a complicated interplay between three different  families of electron-hole transitions: intra-band transitions (i.e.~$\lambda_1=\lambda_2=\lambda_3=\lambda_4$), inter-band transitions (i.e.~$\lambda_1=-\lambda_2=\lambda_3=-\lambda_4$) 
and ``hybrid'' transitions (e.g.~$\lambda_1=\lambda_2=\lambda_3=-\lambda_4$). 
The latter ones are of course absent in the first-order tensor ${\bm \sigma}^{(1)}$.
In total, there are $12$ contributions of ``hybrid'' nature, resulting from both intra- and inter-band processes along the fermion loop in Fig.~\ref{fig:one}. 

We are now in the position to take the uniform $|{\bm q}|\to 0$ limit by expanding Eqs.~(\ref{eq:pi32})-(\ref{eq:form_factor}) in powers of ${\bm q}$. We hasten to emphasize that the expansion of the form factor 
$F_{\lambda_1\dots\lambda_4}({\bm k},{\bm q},{\bm q},{\bm q})$ up to third order in ${\bm q}$ cannot be obtained by simply expanding each factor in Eq.~(\ref{eq:form_factor}) up to linear order in ${\bm q}$---cf.~Eqs.~(16) and~(22) in Ref.~\onlinecite{mikhailov_arxiv_2015}. 

After lengthy but straightforward calculations, we obtain the desired result for THG generation in graphene:
\begin{equation}\label{eq:sigma3}
\sigma^{(3)}_{xxxx}= {\widetilde \sigma}_{xxxx; 1} + {\widetilde \sigma}_{xxxx; 2} + \sigma_{xxxx; {\rm FS}}~.
\end{equation}
The last term on the right-hand side of Eq.~(\ref{eq:sigma3}), $\sigma_{xxxx; {\rm FS}}$, 
is a Fermi surface contribution, which is controlled by an integral over energy whose 
integrand is pinned at the Fermi surface by the first, second, and third derivatives of $n_{\rm F}(E)$:
\begin{eqnarray}\label{eq:FS_contribution}
\sigma_{xxxx; {\rm FS}}&=&
i \kappa \int^\infty_0 dE
\Big\{
\big[n^{\prime}_{\rm F}(E)+n^{\prime}_{\rm F}(-E) \big] f(E) \nonumber\\
&+&\big[n^{\prime\prime}_{\rm F}(E)-n^{\prime\prime}_{\rm F}(-E) \big] g(E)\nonumber\\
&+&\big [n^{\prime\prime\prime}_{\rm F}(E)+n^{\prime\prime\prime}_{\rm F}(-E) \big] h(E)
\Big \}~,
\end{eqnarray}
where $\kappa=N_{\rm f}e^4 \hbar v^2_{\rm F}/ (32\pi)$, $n^{\prime}_{\rm F}(E)$ is shorthand for the derivative $dn_{\rm F}(E)/dE$, and $n^{\prime}_{\rm F}(-E)$ is a shorthand for $n^{\prime}_{\rm F}(E)|_{E \to -E}$. (Similar shorthands have been used for the second and third derivatives.)
Explicit expressions for the functions $f(E)$, $g(E)$, and $h(E)$ are reported in the Appendix.
The terms ${\widetilde \sigma}_{xxxx; 1}$ and 
${\widetilde \sigma}_{xxxx; 2}$ are defined by:
\begin{equation}\label{eq:sigmatilde}
{\widetilde \sigma}_{xxxx; 1,2} = i  \kappa
\int^\infty_0 d E
\left [  n_{\rm F}(E)-n_{\rm F}(-E) \right ] F_{1,2}(E)~,
\end{equation}
where
\begin{eqnarray}\label{eq:powerlaw}
F_{1}(E) &=&
\Bigg\{
\frac{1}{E^2(\hbar\omega_{+})^3} 
+\frac{4}{(\hbar\omega_{+})^2} \bigg [\frac{1} {(\hbar\omega_{+} +2E)^3} \nonumber\\
&+&\frac{1} {(\hbar\omega_{+} - 2E)^3} \bigg]- 
\frac{8}{(\hbar\omega_{+})^3} \bigg[\frac{1} {(\hbar\omega_{+} + 2E)^2} \nonumber \\
&+&\frac{1} {(\hbar\omega_{+} - 2 E)^2} \bigg]
+\frac{2}{(\hbar\omega_{+})^3} \bigg[\frac{1} {(\hbar\omega_{+} + E)^2} \nonumber \\
&+&\frac{1} {(\hbar\omega_{+} - E)^2} \bigg]
\Bigg \}
\end{eqnarray}
and
\begin{eqnarray}\label{eq:log}
F_{2}(E) &=&
\Bigg \{-\frac{8}{3(\hbar\omega_{+})^4} \bigg [\frac{1} {\hbar\omega_{+} + E} + 
\frac{1} {\hbar\omega_{+} - E } \bigg]\nonumber\\
&+&\frac{17}{12(\hbar\omega_{+})^4} \bigg[\frac{1} {\hbar\omega_{+} + 2 E }+\frac{1} {\hbar\omega_{+} - 2 E } \bigg]\nonumber \\ 
&+&
\frac{5}{4(\hbar\omega_{+})^4} \bigg[\frac{1} {\hbar\omega_{+} + 2E/3}+\frac{1} {\hbar\omega_{+} 
- 2E/3} \bigg ]\Bigg\}~.\nonumber\\
\end{eqnarray}
In Eqs.~(\ref{eq:powerlaw})-(\ref{eq:log}) we have introduced the shorthand $\omega_+ \equiv \omega + 
i\eta/\hbar$. Note that, for large $E$, $F_{1}(E)$ decays faster than $1/E$, while $F_{2}(E)$ decays exactly like $1/E$. As a consequence,  
${\widetilde \sigma}_{xxxx; 1}$ 
(${\widetilde \sigma}_{xxxx; 2}$) contains power-laws (logarithms). The explicit calculation of ${\widetilde \sigma}_{xxxx; 1,2}$ does not require an ultraviolet cutoff, which would break gauge invariance~\cite{principi_prb_2009,chirolli_prl_2012,abedinpour_prb_2011}.

We remind the reader that a similar calculation of the first-order diagram in Fig.~\ref{fig:one}(a) yields the well-known result
\begin{eqnarray}\label{eq:first-order-conductivity}
\sigma^{(1)}_{xx}(-\omega;\omega) &=& \frac{i}{\pi} \bigg \{ \frac{{\cal D}}{ \omega_+}
-\sigma_{\rm uni} \int^{\infty}_0dE~\big[n_{\rm F}(E)-n_{\rm F}(-E)\big] 
\nonumber\\
&\times&\bigg[ \frac{2}{\hbar\omega_{+} + 2 E}+\frac{2}{\hbar\omega_{+} -2 E} \bigg] \bigg \}~,
\end{eqnarray}
where ${\cal D} = -4(\sigma_{\rm uni}/\hbar)\int^{\infty}_0dE  ~E[n^\prime_{\rm F}(E)+n^\prime_{\rm F}(-E)]= 8\sigma_{\rm uni}\ln[2\cosh(\beta\mu/2)]/(\beta\hbar)$ is the finite-$T$ Drude weight~\cite{wagner_nanolett_2014}, $\sigma_{\rm uni} = N_{\rm f} e^2/(16\hbar)$ being the universal optical conductivity~\cite{kotov_rmp_2012}. The first term on the right-hand side of Eq.~(\ref{eq:first-order-conductivity}), which is proportional to the Drude weight, 
is an intra-band contribution, while the second term is an inter-band contribution. Due to the form of the integrand in Eq.~(\ref{eq:sigmatilde}) and for its similarity with the integrand in the second term in curly brackets in Eq.~(\ref{eq:first-order-conductivity}), 
we will refer to ${\widetilde \sigma}_{xxxx; 1}$ and ${\widetilde \sigma}_{xxxx; 2}$ as to ``inter-band'' contributions to the third-order conductivity. 

Integrating Eq.~(\ref{eq:FS_contribution}) by parts, it is possible to show (see Appendix) that the following equality holds true:
\begin{equation}\label{eq:intra_interA}
\sigma^{\rm FS}_{xxxx}(-3\omega;\omega,\omega,\omega) = - {\widetilde \sigma}_{xxxx; 1}(-3\omega;\omega,\omega,\omega)~.
\end{equation}
We therefore conclude that $\sigma_{xxxx}(-3\omega;\omega,\omega,\omega) = {\widetilde \sigma}_{xxxx; 2}(-3\omega;\omega,\omega,\omega)$. Eq.~(\ref{eq:intra_interA}) is the most important result of this work and implies the {\it absence} of power-law terms 
in the final result for $\sigma_{xxxx}(-3\omega;\omega,\omega,\omega)$.

In the $T =0$ limit, we find
\begin{eqnarray}\label{eq:final_result}
&&\sigma^{(3)}_{xxxx} (-3\omega; \omega,\omega,\omega) =\frac{i \kappa}{24 (\hbar\omega_{+})^4}
\big[
17 {\cal G}_{\eta}(\hbar\omega, 2|E_{\rm F}|) 
\nonumber\\&&
- 64 {\cal G}_{\eta}(\hbar\omega,|E_{\rm F}|) 
+ 45{\cal G}_{\eta}(\hbar\omega,2|E_{\rm F}|/3) \big]
\end{eqnarray}
where ${\cal G}_{\eta}(\hbar\omega,E) = \ln[(E + \hbar\omega_+)/(E- \hbar\omega_{+})]$. The final result for THG can be obtained by taking the limit $\eta \to 0^+$ in Eq.~(\ref{eq:final_result}), with 
${\cal G}_{\eta\to 0^+}(\hbar\omega,E) =  \ln|(E+\hbar\omega)/(E-\hbar\omega)| +i\pi \Theta(|\hbar\omega|-E)$. Finally, we observe that Eq.~(\ref{eq:final_result}) is well behaved in the undoped $E_{\rm F} \to 0$ limit, i.e.
\begin{equation}\label{eq:undoped_Tzero}
\lim_{E_{\rm F}\to 0}\lim_{\eta \to 0^+}\sigma^{(3)}_{xxxx}(-3\omega; \omega,\omega,\omega) 
= \frac{\pi \kappa}{12(\hbar\omega)^4}~.
\end{equation}
On the contrary, the final result in Eqs.~(10)-(12) of Ref.~\onlinecite{mikhailov_prb_2014} is ill defined in the $E_{\rm F} \to 0$ limit.

Since the third-order conductivity is known analytically at $T=0$, finite-$T$ effects are most conveniently studied by using the Maldague identity~\cite{Giuliani_and_Vignale}. In our case, this yields the following integral representation for the third-order conductivity at $T\neq 0$:
\begin{equation}
\sigma^{(3)}_{xxxx}|_{T\neq 0} =\beta\int^{\infty}_{-\infty}dE~\frac{ \sigma^{(3)}_{xxxx}|_{\{T=0,~E_{\rm F} \to E\}}  }{4 \cosh^2(\beta (E-\mu)/2)}~, 
\end{equation}
where the $T=0$ result $\sigma^{(3)}_{xxxx}|_{\{T=0,~E_{\rm F} \to E\}}$ can be obtained from Eq.~(\ref{eq:final_result}). The chemical potential as a function of $T$ can be found by solving $\beta^2E^2_{\rm F} =2| {\rm Li}_2[-\exp (-\beta\mu)]  -  {\rm Li}_2[-\exp(\beta\mu)]|$, where ${\rm Li}_{2}[x]$ is the dilogarithm function. In the limit $\eta\to 0^+$, the real part of $\sigma^{(3)}_{xxxx}$ can be written in a closed form for any value of $T$ and $E_{\rm F}$: 
\begin{eqnarray}\label{eq:finite_T}
{\rm Re}[\sigma^{(3)}_{xxxx}]
&=&  \frac{\pi \kappa}{24(\hbar\omega)^4}[2 + 17 n_{\rm F}(\hbar\omega/2)
- 64n_{\rm F}(\hbar\omega)\nonumber\\
&+&45n_{\rm F}(3\hbar\omega/2)]~.
\end{eqnarray}
We were not able to find a similar analytical expression for ${\rm Im}[\sigma^{(3)}_{xxxx}]$ at arbitrary $T$. For an undoped system, $\mu=0$ at any $T$ and ${\rm Im}[\sigma^{(3)}_{xxxx}]=0$.  

At this stage, one may be tempted to take into account disorder by introducing a phenomenological relaxation time $\tau$ through the replacement $\eta \to \hbar/\tau$ or $\omega_{+}\to \omega + i/\tau$. In general, this procedure yields a ``non-conserving'' approximation~\cite{kadanoff-baym} for optical and transport response functions. The case of the ordinary density-density response function is discussed in Refs.~\onlinecite{Giuliani_and_Vignale,mermin_prb_1970}. 
In the diagrammatic language, this replacement takes into account disorder-induced self-energy corrections to response functions, while conserving approximations require to treat on an equal footing self-energy {\it and} vertex corrections. This is why in this work we present results for the clean system and postpone the analysis of THG in disordered graphene sheets to future work.

\begin{figure}[t]
\centering
\includegraphics[width=1.0\linewidth]{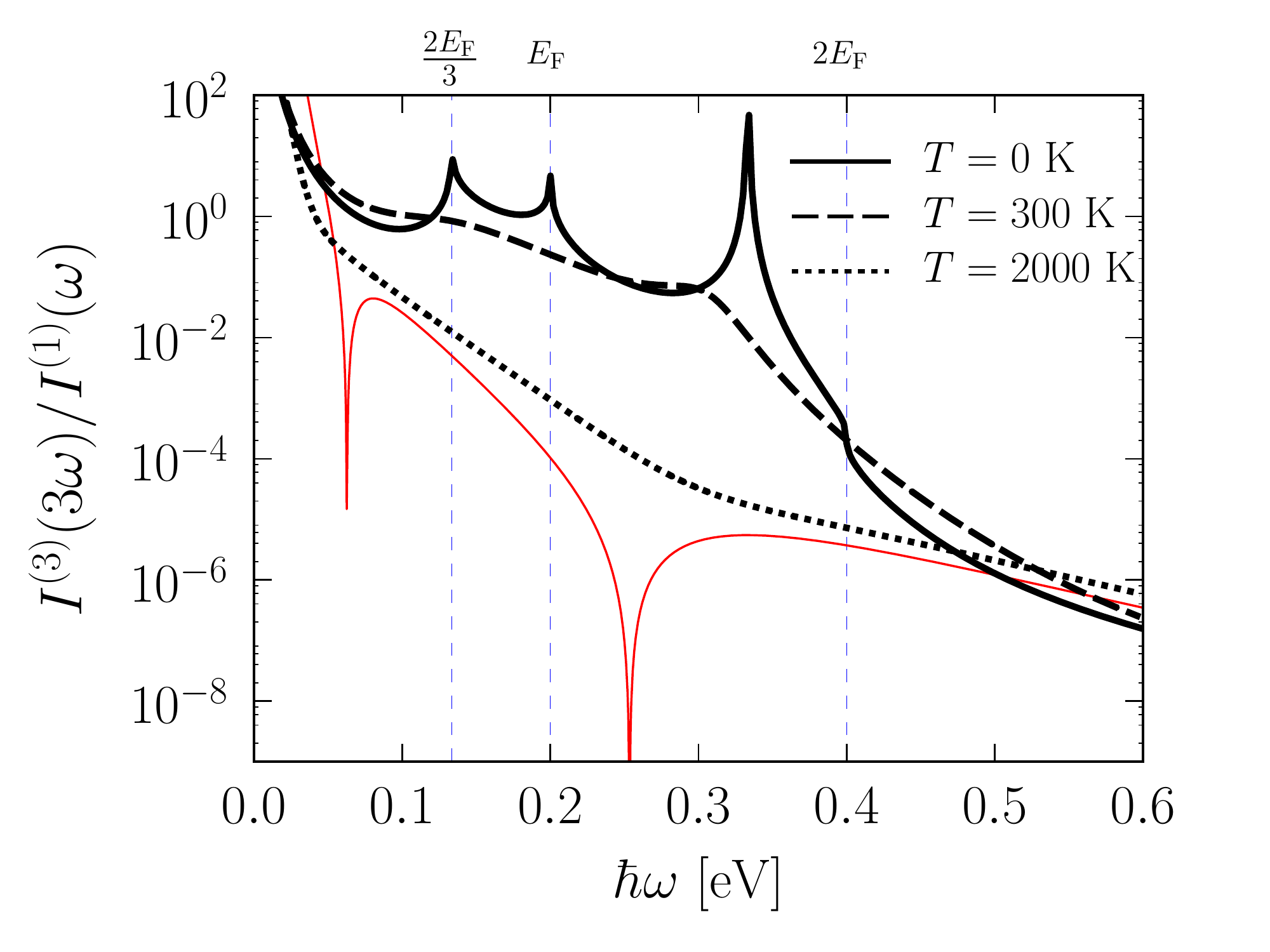}
\caption{(Color online) The ratio $I^{(3)}(3\omega)/I^{(1)}(\omega)$ for a clean graphene sheet is plotted as a function of the incident photon energy $\hbar\omega$ (in ${\rm eV}$) and for an incident power $I_{\rm i} =1~{\rm GW}/{\rm cm}^2$. Different curves refer to different values of temperature $T$. Results in this figure have been obtained by setting $E_{\rm F}=200~{\rm meV}$. The solid thin red line represents the result for undoped graphene at $T=2000$~K, as from Eq.~(\ref{eq:finite_T}) and $\mu=0$. The dashed vertical lines label the three special photon frequencies: $\hbar\omega=2E_{\rm F}/3$, $E_{\rm F}$, and $2E_{\rm F}$. \label{fig:two}}
\end{figure}

{\it Up-conversion efficiency.---}For the sake of completeness, we finally present our predictions for the efficiency of the THG process in a clean graphene sheet at finite $T$. Using the relation ${\bm J}^{(n)}(\omega)=-i  \omega_{+}{\bm P}^{(n)}(\omega)$ between the Fourier transforms of the induced current and polarization ${\bm P}^{(n)}(t)$, we find
$P^{(3)}_{\ell}(\omega_\Sigma) = (i/\omega_{\Sigma,+}) \sum_{\alpha_1\dots \alpha_3}\sigma^{(3)}_{\ell \alpha_1\dots\alpha_3}\Pi^{3}_{i=1}E_{\alpha_i}(\omega_i)$, where $\omega_{\Sigma,+} \equiv \omega_{\Sigma}+i\eta/\hbar$. Assuming a monochromatic incident light beam, linearly polarized along the $\hat{\bm x}$ direction, we find the following expression for the THG intensity, $I^{(3)}(3\omega)$, in units of the first-order intensity $I^{(1)}(\omega)$:
\begin{equation}
\frac{I^{(3)}(3\omega)}{I^{(1)}(\omega)} \equiv \left | \frac{P^{(3)}_x(3\omega)}{P^{(1)}_x(\omega)} \right |^2 = \left [\frac{2 I_{\rm i}}{3 n_{\rm r} \epsilon_{0} c} \right ]^2
 \left | \frac{{\sigma}^{(3)}_{xxxx}}{{\sigma}^{(1)}_{xx}} \right |^2~.
\end{equation}
Here, $I_{\rm i}=n_{\rm r} \epsilon_{0} c |{\bm E}|^2/2$ is the intensity of incident light where $|{\bm E}|$ is the time average of the incident electric field, $\epsilon_{0} \simeq 8.85 \times 10^{-12}~{\rm C}/({\rm V m})$ is the vacuum permittivity, $c \simeq 3\times 10^8~{\rm m}/{\rm s}$, and $n_{\rm r} \simeq 1$. The quantity $I^{(3)}(3\omega)/I^{(1)}(\omega)$ is shown in Fig.~\ref{fig:two} for an incident power $I_{\rm i} =1~{\rm GW}/{\rm cm}^2$. (This is the peak power used in on-going experiments on THG in doped graphene sheets.) According to Eq.~(\ref{eq:final_result}), there are three logarithmic divergences at photon energies $\hbar\omega= 2E_{\rm F}/3$, $E_{\rm F}$, and $2E_{\rm F}$ in the $T=0$ expression of ${\sigma}^{(3)}_{xxxx}$ (marked by vertical dashed lines in the figure). Finite-$T$ effects rapidly smooth these weak singularities out. The main peak occurs at $\hbar\omega=2E_{\rm F}/3$. The peak at $\hbar\omega \simeq 1.667~E_{\rm F}$ is due to the fact that intra- and inter-band contributions to the first-order conductivity cancel out~\cite{mikhailov_prb_2014,mikhailov_arxiv_2015} at this photon frequency and at $T=0$, yielding $|\sigma^{(1)}_{xx}(-\omega;\omega)| = |{\rm Im}[\sigma^{(1)}_{xx}(-\omega;\omega)]|=0$. (We remind the reader that ${\rm Re}[\sigma^{(1)}_{xx}]=0$ in the single-particle optical gap $\hbar\omega < 2 E_{\rm F}$.) The undoped result at finite $T$ shows two sharp structures at photon energies $\hbar\omega \simeq 0.362/\beta$ and $\hbar\omega \simeq 1.462/\beta$, which correspond to solutions of the equation $|\sigma^{(3)}_{xxxx}(-3\omega;\omega,\omega,\omega)|=0$ in the undoped case, as it can be readily checked by utilizing Eq.~(\ref{eq:finite_T}). For illustration purposes, in Fig.~\ref{fig:two} we show the undoped result at $T =2000~{\rm K}$ (solid red line).

In summary, we have presented a diagrammatic theory of THG in doped graphene. We have carried out explicit calculations in the scalar potential gauge, discovering an exact cancellation between the Fermi surface contribution to the third-order conductivity and power-law contributions of inter-band nature. Only logarithms survive in the final result, Eq.~(\ref{eq:final_result}). We believe that our results shed light on the on-going discussion~\cite{mikhailov_prb_2014,mikhailov_arxiv_2015,cheng_njp_2014,cheng_prb_2015} about THG in graphene, showing that the lack of consensus is due to the anomalous cancellation discussed in this work. Calculations of THG can also be carried out in the vector potential gauge, but in this case, for the reasons mentioned above, it is safer to use lattice Hamiltonians rather than the MDF model---see e.g.~Ref.~\onlinecite{wehling_prb_2015} for the case of second harmonic generation in the presence of broken inversion symmetry.

{\it Acknowledgements.---}We thank G. Cerullo, A.C. Ferrari, S.A. Jafari, and M.I. Katsnelson for many useful discussions. This work was supported by the EC under the Graphene Flagship program (contract no.~CNECT-ICT-604391).

\appendix

\section{Fermi surface contribution and proof of Eq.~(15) in the main text}
In Eq.~(10) of the main text we have defined the following functions:
\begin{eqnarray}
f(E) &=& 
-\frac{1}{E (\hbar\omega_{+})^3} 
-\frac{2}{E^3 \hbar\omega_{+}} 
-\frac{1}{E^3 (\hbar\omega_{+} +2E)}\nonumber\\ 
&-&\frac{1}{E^3 (\hbar\omega_{+} - 2E)}
-\frac{1}{2 E^2 (\hbar\omega_{+} + 2E)^2} 
\nonumber\\
&+& \frac{1}{2 E^2 (\hbar\omega_{+} - 2E)^2}
+\frac{2}{E^3 (\hbar\omega_{+} +E)} 
\nonumber\\
&+&\frac{2}{E^3 (\hbar\omega_{+} - E)}~,
\end{eqnarray}
\begin{eqnarray}
g(E) &=&
\frac{1}{(\hbar\omega_{+})^3} 
- \frac{1}{4 E^2 \hbar\omega_{+}}
+ \frac{1}{8 E^2 (\hbar\omega_{+} + 2 E)} 
\nonumber\\
&+& \frac{1}{8 E^2 (\hbar\omega_{+} - 2 E)}~,
\end{eqnarray}
and
\begin{equation}
h(E)=\frac{E}{(\hbar\omega_{+})^3}~.
\end{equation}

We now briefly summarize the steps that lead to Eq.~(15) in the main text. We start by integrating by parts the Fermi surface term that contains the first derivative of the Fermi-Dirac distribution function: 
\begin{eqnarray}\label{eq:bypart1}
&&\int^\infty_0 dE~\left [n^{\prime}_{\rm F}(E)+n^{\prime}_{\rm F}(-E) \right ] f(E) =\nonumber\\
&-&\int^\infty_0 dE~\left [n_{\rm F}(E)-n_{\rm F}(-E) \right ] f^{\prime}(E) \nonumber\\
&+&\Big|f(E)\left[ n_{\rm F}(E)-n_{\rm F}(-E) \right ]\Big|^\infty_0~.
\end{eqnarray}
For the boundary term, we have used the shorthand
\begin{equation}
\Big|{\cal O}(E)\Big|^{b}_{a} \equiv \lim_{E \to b}  {\cal O}(E) - \lim_{E \to a} {\cal O}(E)~.
\end{equation}
Integrating twice the Fermi surface term containing the second derivative of the Fermi-Dirac distribution function we find:
\begin{eqnarray}\label{eq:bypart2}
&&\int^\infty_0 dE~\left [n^{\prime\prime}_{\rm F}(E)-n^{\prime\prime}_{\rm F}(-E) \right ] g(E) =\nonumber\\
&&\int^\infty_0 dE~\left [n_{\rm F}(E)-n_{\rm F}(-E) \right ] g^{\prime\prime}(E) 
\nonumber\\
&+&
\Big|
g(E) \left [n^{\prime}_{\rm F}(E)+n^{\prime}_{\rm F}(-E) \right ] 
\nonumber\\
&-&g^{\prime}(E)\left [n_{\rm F}(E)-n_{\rm F}(-E) \right ]  
\Big|^\infty_0~.
\end{eqnarray} 
Similarly, we need to carry three integrations by parts in the Fermi surface term containing the third derivative of the Fermi-Dirac distribution function:
\begin{eqnarray}\label{eq:bypart3}
&&\int^\infty_0 dE~\left [n^{\prime\prime\prime}_{\rm F}(E)+n^{\prime\prime\prime}_{\rm F}(-E) \right ] h(E) =\nonumber\\
&-&\int^\infty_0 dE \left [n_{\rm F}(E)-n_{\rm F}(-E) \right ] h^{\prime\prime\prime}(E) 
\nonumber\\
&+&\Big| 
\left [n^{\prime\prime}_{\rm F}(E)-n^{\prime\prime}_{\rm F}(-E) \right ] h(E) 
\nonumber\\&-&
\left [n^{\prime}_{\rm F}(E)+n^{\prime}_{\rm F}(-E) \right ] h^{\prime}(E) \nonumber\\
&+&\left [n_{\rm F}(E)-n_{\rm F}(-E) \right ] h^{\prime\prime}(E) 
\Big|^\infty_0~.
\end{eqnarray} 
By using Eqs.~(\ref{eq:bypart1})-(\ref{eq:bypart3}) we find the following result: 
\begin{eqnarray}\label{eq:with_boundary_terms}
&&\sigma^{\rm FS}_{xxxx}  (-3\omega;\omega,\omega,\omega)=
-i\kappa \int^\infty_0 dE~\left [n_{\rm F}(E)-n_{\rm F}(-E) \right]  \nonumber\\
&\times& \left [ f^{\prime}(E) -g^{\prime\prime}(E) + h^{\prime\prime\prime}(E)\right ]
\nonumber\\&+&
i\kappa 
\Big|
\left [n_{\rm F}(E)-n_{\rm F}(-E) \right ] \left [ f(E)-g^\prime(E)+h^{\prime\prime}(E) \right ]
\nonumber\\
&+&\left [n^\prime_{\rm F}(E)+n^\prime_{\rm F}(-E) \right ] \left[g(E) -h^\prime(E)\right]
\nonumber\\
&+&\left[n^{\prime\prime}_{\rm F}(E)-n^{\prime\prime}_{\rm F}(-E) \right] h(E)
\Big|^\infty_0~.
\end{eqnarray}
For the evaluation of the boundary terms we need the following results:
\begin{eqnarray}
&&\lim_{E \to \infty } \left [n_{\rm F}(E)-n_{\rm F}(-E) \right ] \left [ f(E)-g^\prime(E)+h^{\prime\prime}(E) \right ] = 0~,
\nonumber\\&&
\lim_{E \to 0 } \left [n_{\rm F}(E)-n_{\rm F}(-E) \right ] \left [ f(E)-g^\prime(E)+h^{\prime\prime}(E) \right ] = - A~,
\nonumber\\&&
\lim_{E \to \infty } \left [n^\prime_{\rm F}(E)+n^\prime_{\rm F}(-E) \right ] \left [  g(E) -h^\prime(E)\right ] =0~,
\nonumber\\&&
\lim_{E \to 0 }  \left [n^\prime_{\rm F}(E)+n^\prime_{\rm F}(-E) \right ] \left [  g(E) -h^\prime(E)\right ]  = +A~,
\nonumber \\&&
\lim_{E \to \infty } \left [n^{\prime\prime}_{\rm F}(E)-n^{\prime\prime}_{\rm F}(-E) \right ] h(E) =0~,
\nonumber\\&&
\lim_{E \to 0 } \left [n^{\prime\prime}_{\rm F}(E)-n^{\prime\prime}_{\rm F}(-E) \right ] h(E) =0~,
\end{eqnarray}
where we have introduced $A = 2 n^\prime_{\rm F}(0)/(\hbar\omega_{+})^3$.
We therefore conclude that boundary terms in Eq.~(\ref{eq:with_boundary_terms}) add up to zero. We therefore find 
\begin{eqnarray}
&&\sigma^{\rm FS}_{xxxx}  (-3\omega;\omega,\omega,\omega) =
-i \kappa  \int^\infty_0  dE  \left [n_{\rm F}(E)-n_{\rm F}(-E) \right ] \nonumber\\
&\times& \left [ f^{\prime}(E) -g^{\prime\prime}(E) + h^{\prime\prime\prime}(E)\right ]~.
\end{eqnarray}
Finally, it is easy to show that 
\begin{equation}
 f^{\prime}(E) -g^{\prime\prime}(E) + h^{\prime\prime\prime}(E) = F_1(E)~.
\end{equation}
\end{document}